# Improved Randomized Response Technique for Two Sensitive Attributes

Ewemooje, O.S[*1] and Amahia, G.N[2]

[1]Department of Statistics, Federal University of Technology, Akure, Nigeria.
[2]Department of Statistics, University of Ibadan, Ibadan, Nigeria.
[*] osewemooje@futa.edu.ng; sunsege@yahoo.com;

We proposed new and more efficient estimators for estimating population proportion of respondents belonging to two related sensitive attributes in survey sampling by extending the work of Mangat (1994). Our proposed estimators are more efficient than Lee et al (2013) simple and crossed model estimators as the population proportion of possessing the sensitive attribute increases.

## 1. Introduction

Reliability of data is compromised when sensitive topics on embarrassing or illegal acts such as drunk driving, abortion, alcoholism, illicit drugs usage, tax evasion, illegal possession of arms are required in direct method of data collection in sample survey. Of a major concern is the impact of the response distortion on the survey or test results. Surveys on human population have established the fact that the direct question about sensitive characters often result in either refusal to respond or falsification of the answer (Sidhu et al, 2009). However, obtaining valid and reliable information is a prerequisite for obtaining meaningful results. Hence, there is need to ensure confidentiality of respondents which will in-turn lead to more reliable information.

Warner (1965) developed an interviewing procedure designed to reduce or eliminate this bias and called it Randomized Response Technique (RRT). This allows researchers to obtain sensitive information while guaranteeing respondents' privacy. Many works have been done to improve the strategy introduced by Warner (1965), see Bouza et al (2010) for review.

Mangat (1994) proposed a strategy in which respondent is instructed to say 'yes' if he/she belong to an attribute A. if not, he/she is required to use the Warner randomized device consisting of two statements:

- I belong to attribute A with probability P
- I do not have attribute A with probability 1 - P

The probability of a 'yes' answer for the procedure is given by:

$$\alpha = \pi + (1 - \pi)(1 - P) \tag{1.1}$$

where $\pi$ is the population proportion of respondents possessing the sensitive attribute.

The proposed Mangat unbiased estimator for $\pi$ is:

$$\hat{\pi}_M = \frac{\hat{\alpha} - 1 + P}{P} \tag{1.2}$$

where $\hat{\alpha}$ is the observed proportion of 'yes' answers obtained from sampled individuals. The variance of the estimator is given as:

$$V(\hat{\pi}_M) = \frac{\hat{\alpha}(1 - \hat{\alpha})}{(n-1)P^2} \tag{1.3}$$

where n is the sample size

In order to improve the work of Warner (1965), Odumade and Singh (2009) suggested a new randomized model using two decks of cards. Each of these decks of cards is the same as in Warner model with varied probabilities. All the respondents were made to go through the device twice for a single attribute. Other works in this regards include Arnab et al (2012), Singh and Sedory (2012), Song and Kim (2012).

It is important to note that Warner's and others' designs can only capture one attribute at a time and many potential information were lost. This is a limitation, as no researcher sets out to collect only one item in a survey but multiple items. Therefore, the research work focus on providing a models for measuring hidden respondent characteristics on two sensitive items and their interaction. Also, investigate the performance of the developed strategies.

Christofides (2005) developed a new strategy of estimating the proportion of individuals having two sensitive characteristics at the same time. Lee et al (2013) proposed a simple model which is a special case of Christofides (2005) with probabilities of (yes,yes), (yes,no), (no,yes) and (no,no) as denoted by $\theta_{11}, \theta_{10}, \theta_{01}$ and $\theta_{00}$ respectively, where;

$$\theta_{11} = (2P-1)(2T-1)\pi_{AB} + (2P-1)(1-T)\pi_A + (1-P)(2T-1)\pi_B + (1-P)(1-T) \quad (1.4)$$

$$\theta_{10} = -(2P-1)(2T-1)\pi_{AB} + (2P-1)T\pi_A + (1-P)(2T-1)\pi_B + (1-P)T \quad (1.5)$$

$$\theta_{01} = -(2P-1)(2T-1)\pi_{AB} + (2P-1)(1-T)\pi_A + P(2T-1)\pi_B + P(1-T) \quad (1.6)$$

$$\theta_{00} = (2P-1)(2T-1)\pi_{AB} + (2P-1)T\pi_A + P(2T-1)\pi_B + PT \quad (1.7)$$

The unbiased estimators of the proportion $\pi_A, \pi_B$ and $\pi_{AB}$ are given by:

$$\hat{\pi}_{A(SM)} = \frac{\hat{\theta}_{11} + \hat{\theta}_{10} - \hat{\theta}_{01} - \hat{\theta}_{00} + (2P-1)}{2(2P-1)} \quad (1.8)$$

$$\hat{\pi}_{B(SM)} = \frac{\hat{\theta}_{11} - \hat{\theta}_{10} + \hat{\theta}_{01} - \hat{\theta}_{00} + (2T-1)}{2(2T-1)} \quad (1.9)$$

$$\hat{\pi}_{AB(SM)} = \frac{(P+T)\hat{\theta}_{11} + (T-P)\hat{\theta}_{10} + (P-T)\hat{\theta}_{01} + (2-P-T)\hat{\theta}_{00} - T(1-P) - P(1-T)}{2(2P-1)(2T-1)} \quad (1.10)$$

For $T \neq 0.5$ and $P \neq 0.5$ where $\hat{\theta}_{11} = n_{11}/n$, $\hat{\theta}_{10} = n_{10}/n$, $\hat{\theta}_{01} = n_{01}/n$ and $\hat{\theta}_{00} = n_{00}/n$. The variances of the estimators are given as:

$$V(\hat{\pi}_{A(SM)}) = \frac{\pi_A(1-\pi_A)}{n} + \frac{P(1-P)}{n(2P-1)^2} \quad (1.11)$$

$$V(\hat{\pi}_{B(SM)}) = \frac{\pi_B(1-\pi_B)}{n} + \frac{T(1-T)}{n(2T-1)^2} \quad (1.12)$$

$$V(\hat{\pi}_{AB(SM)}) = \frac{\pi_{AB}(1-\pi_{AB})}{n} + \frac{(2P-1)^2 T(1-T)\pi_A + P(1-P)(2T-1)^2 \pi_B + PT(1-P)(1-T)}{n(2P-1)^2(2T-1)^2} \quad (1.13)$$

for $T = P \neq 0.5$

Lee et al (2013) also proposed new model called "Crossed Model". This they established to be more efficient than the simple model. Perri et al (2015) applied the crossed model in estimating induced abortion and foreign irregular presence in Calabria, Italy.

The unbiased estimators of the population proportion $\pi_A, \pi_B$ and $\pi_{AB}$ for the mixed model are given by:

$$\hat{\pi}_{A(CM)} = \frac{1}{2} + \frac{(T-P+1)(\hat{\theta}_{11}^* - \hat{\theta}_{00}^*) + (P+T-1)(\hat{\theta}_{10}^* - \hat{\theta}_{01}^*)}{2(P+T-1)} \quad (1.14)$$

$$\hat{\pi}_{B(CM)} = \frac{1}{2} + \frac{(T-P+1)(\hat{\theta}_{11}^* - \hat{\theta}_{00}^*) + (P+T-1)(\hat{\theta}_{01}^* - \hat{\theta}_{10}^*)}{2(P+T-1)} \quad (1.15)$$

$$\hat{\pi}_{AB(CM)} = \frac{PT\hat{\theta}_{11}^* - (1-P)(1-T)\hat{\theta}_{00}^*}{\{PT+(1-P)(1-T)\}(P+T-1)} \qquad (1.16)$$

For P + T ≠ 1

The variances of these estimators are given as:

$$V(\hat{\pi}_{A(CM)}) = \frac{\pi_A(1-\pi_A)}{n} + \frac{(1-P)[T\{PT+(1-P)(1-T)\}(1-\pi_A-\pi_B+2\pi_{AB})]}{n(P+T-1)^2} \qquad (1.17)$$

$$V(\hat{\pi}_{B(CM)}) = \frac{\pi_B(1-\pi_B)}{n} + \frac{(1-T)[P\{PT+(1-P)(1-T)\}(1-\pi_A-\pi_B+2\pi_{AB})]}{n(P+T-1)^2} \qquad (1.18)$$

$$V(\hat{\pi}_{AB(CM)}) = \frac{\pi_{AB}(1-\pi_{AB})}{n} + \frac{1}{n\{PT+(1-P)(1-T)\}(P+T-1)^2}[\pi_{AB}\{P^2T^2+(1-P)^2(1-T)^2 -$$
$$\{PT+(1-P)(1-T)\}(P+T-1)^2\} + PT(1-P)(1-T)(1-\pi_A-\pi_B)] \qquad (1.19)$$

## 2.    Propose Design

We consider selecting a sample from a finite population using simple random sample with replacement. Two related sensitivity questions "A" and "B" are posted at each respondent in order to estimate proportion of respondents belonging to character "A" or "B" or "AB". Let population proportion of respondents belonging to character A, B or AB be $\pi_A, \pi_B$ and $\pi_{AB}$ respectively. A procedure similar to that of Mangat(1994) strategy is being presented in which respondent is instructed to answer "yes" if he or she belong to attribute/character "A" if not, he or she is required to draw a card from deck I of cards containing two statements:

- ➢ I belong to character A with probability P
- ➢ I do not belong to character A with probability 1 – P

And answer "yes" or "no" accordingly without reporting the statement on the card to the interviewer. Also, the respondent proceed to next stage by answering "yes" if he or she belong to character "B". If not, he or she is required to draw another card from deck II of cards containing either of the two statements:

- ➢ I belong to character B with probability λ
- ➢ I do not belong to character B with probability 1 – λ

And answer "yes" or "no" accordingly without reporting the statement on the card to the interviewer.

The observed responses can be categorized into four different places: $n_{11}$ (i.e number of respondents that answered "yes" to character A and "yes" to character B), $n_{10}$ (number of respondents that answer "yes" to character A and no to character B), $n_{01}$ (number of respondents that answered "no" to character A and "yes" to character B) and $n_{00}$ (number of respondents that answer "no" to character A and "no" to character B) where $n_{11} + n_{10} + n_{01} + n_{00} = n$ i.e $\sum_{i=0}^{1}\sum_{j=0}^{1} n_{ij} = n$. Therefore we represent the probabilities of (yes, yes), (yes, no), (no, yes), and (no,no) with as $\theta_{11}, \theta_{10}, \theta_{01}$ and $\theta_{00}$ respectively where $\theta_{11} + \theta_{10} + \theta_{01} + \theta_{00} = n$ i.e $\sum_{i=0}^{1}\sum_{j=0}^{1} \theta_{ij} = 1$

Using the propose procedure, we have:

$$\theta_{11} = \alpha_1 \pi_{AB} + \alpha_2 \pi_A + \alpha_3 \pi_B + \alpha_4 \tag{2.1}$$

$$\theta_{10} = -\alpha_1 \pi_{AB} + \alpha_1 \pi_A - \alpha_3 \pi_B + \alpha_3 \tag{2.2}$$

$$\theta_{01} = -\alpha_1 \pi_{AB} - \alpha_2 \pi_A + \alpha_1 \pi_B + \alpha_2 \tag{2.3}$$

$$\theta_{00} = \alpha_1 \pi_{AB} - \alpha_1 \pi_A - \alpha_1 \pi_B + \alpha_1 \tag{2.4}$$

Where: $\alpha_1 = P\lambda$, $\alpha_2 = P(1-\lambda)$, $\alpha_3 = (1-P)\lambda$ and $\alpha_4 = (1-P)(1-\lambda)$

The distance between the observed and the true proportion is minimized using the following expression:

$$\varphi = \frac{1}{2}\sum_{i=0}^{1}\sum_{j=0}^{1}\left(\theta_{ij} - \hat{\theta}_{ij}\right)^2 \tag{2.5}$$

We further differentiate $\varphi$ with respect to $\pi_A, \pi_B$ and $\pi_{AB}$ and equate to zero. Then solve simultaneously in order to obtain unbiased estimators of $\pi_A, \pi_B$ and $\pi_{AB}$.

**Theorem 1:** The proposed unbiased estimators of $\pi_A, \pi_B$ and $\pi_{AB}$ are given by:

$$\hat{\pi}_{A(EA)} = \frac{\hat{\theta}_{11} + \hat{\theta}_{10} - \hat{\theta}_{01} - \hat{\theta}_{00} + (2P-1)}{2P} \tag{2.6}$$

$$\hat{\pi}_{B(EA)} = \frac{\hat{\theta}_{11} - \hat{\theta}_{10} + \hat{\theta}_{01} - \hat{\theta}_{00} + (2\lambda-1)}{2\lambda} \tag{2.7}$$

$$\hat{\pi}_{AB(EA)} = \frac{(2P+2\lambda-1)\hat{\theta}_{11} - (2P-2\lambda+1)\hat{\theta}_{10} + (2P-2\lambda-1)\hat{\theta}_{01} - (2P+2\lambda-3)\hat{\theta}_{00} + (2P-1)(2\lambda-1)}{4P\lambda} \tag{2.8}$$

For $P > 0$ and $\lambda > 0$.

The variance of $\pi_A, \pi_B$ and $\pi_{AB}$ can be obtained by:

$$V\left(\hat{\pi}_{A(EA)}\right) = \left\{\frac{\hat{\theta}_{11} + \hat{\theta}_{10} - \hat{\theta}_{01} - \hat{\theta}_{00} + (2P-1)}{2P}\right\}$$

$$V\left(\hat{\pi}_{B(EA)}\right) = \left\{\frac{\hat{\theta}_{11} - \hat{\theta}_{10} + \hat{\theta}_{01} - \hat{\theta}_{00} + (2\lambda - 1)}{2\lambda}\right\}$$

$$V\left(\hat{\pi}_{AB(EA)}\right) = \left\{\frac{(2P + 2\lambda - 1)\hat{\theta}_{11} - (2P - 2\lambda + 1)\hat{\theta}_{10} + (2P - 2\lambda - 1)\hat{\theta}_{01} - (2P + 2\lambda - 3)\hat{\theta}_{00} + (2P - 1)(2\lambda - 1)}{4P\lambda}\right\}$$

**Lemma:** The following variance and covariance operators are used in obtaining the expression for variances of the unbiased estimators;

$$V(x_{11}) = \theta_{11}(1 - \theta_{11}), \quad V(x_{10}) = \theta_{10}(1 - \theta_{10})$$
$$V(x_{01}) = \theta_{01}(1 - \theta_{01}), \quad V(x_{00}) = \theta_{00}(1 - \theta_{00})$$
$$C(x_{11}, x_{10}) = \theta_{11}\theta_{10}, \quad C(x_{11}, x_{01}) = \theta_{11}\theta_{01}, \quad C(x_{11}, x_{00}) = \theta_{11}\theta_{00}$$
$$C(x_{10}, x_{01}) = \theta_{10}\theta_{01}, \quad C(x_{10}, x_{00}) = \theta_{10}\theta_{00}, \quad C(x_{01}, x_{00}) = \theta_{01}\theta_{00}$$

where V and C are the operators of variance and covariance over the randomized response device respectively. And also;

$x_{11}$ is obtained when we have 'yes' response for character A and 'yes' response for character B

$x_{10}$ is obtained when we have 'yes' response for character A and 'no' response for character B

$x_{01}$ is obtained when we have 'no' response for character A and 'yes' response for character B

$x_{00}$ is obtained when we have 'no' response for character A and 'no' response for character B

**Theorem 2**: The variances of the proposed unbiased estimators; $\hat{\pi}_A, \hat{\pi}_B$ and $\hat{\pi}_{AB}$ are given respectively by:

$$V(\hat{\pi}_{A(EA)}) = \frac{\pi_A[(2P-1) - P\pi_A] + (1-P)}{nP} \tag{2.9}$$

$$V(\hat{\pi}_{B(EA)}) = \frac{\pi_B[(2\lambda-1) - \lambda\pi_B] + (1-\lambda)}{n\lambda} \tag{2.10}$$

$$V(\hat{\pi}_{AB(EA)}) = \frac{\pi_{AB}[(2P-1)(2\lambda-1) - P\lambda\pi_{AB}] + (2P-1)(1-\lambda)\pi_A + (1-P)(2\lambda-1)\pi_B + (1-P)(1-\lambda)}{nP\lambda} \tag{2.11}$$

For P > 0 and λ > 0.

## 3. Efficiency Comparison

**Theorem 3:** The propose estimators $\pi_{A(EA)}, \pi_{B(EA)}$ and $\pi_{AB(EA)}$ will be more efficient than estimators $\pi_{A(SM)}, \pi_{B(SM)}$ and $\pi_{AB(SM)}$ which are due to Lee et al respectively if:

i. $\pi_A > \frac{(3P-1)(P-1)}{(2P-1)^2}$ , P ≠ 0.5 \hfill (3.1)

ii. $\pi_B > \frac{(3\lambda-1)(\lambda-1)}{(2\lambda-1)^2}$ , λ ≠ 0.5 \hfill (3.2)

iii. $\pi_{AB} > \dfrac{(2P-1)^2(1-\lambda)\pi_A[(2P-1)(2\lambda-1)^2-P\lambda^2] + (1-P)(2\lambda-1)^2\pi_B[(2P-1)^2(2\lambda-1)-P^2\lambda] + (1-P)(1-\lambda)[(2P-1)^2(2\lambda-1)^2-P^2\lambda^2]}{(2P-1)^2(2\lambda-1)^2[P\lambda-(2P-1)(2\lambda-1)]}$ (3.3)

$P \neq 0.5, \lambda \neq 0.5$

Note: $\pi_{A(EA)}$ is the estimator $\pi_A$ proposed by Ewemooje and Amahia while $\pi_{A(SM)}$ and $\pi_{A(CM)}$ are the simple and crossed model estimators of $\pi_A$ respectively as proposed by Lee, Sedory and Singh. These are applicable to other estimators.

Hence, we obtain the relative efficiency of the propose estimators $\pi_{A(EA)}, \pi_{B(EA)}$ and $\pi_{AB(EA)}$ with respect to the estimators $\pi_{A(SM)}, \pi_{B(SM)}$ and $\pi_{AB(SM)}$ respectively:

$$RE(\hat{\pi}_{A(EA)}, \hat{\pi}_{A(SM)}) = \dfrac{MSE(\hat{\pi}_{A(SM)})}{MSE(\hat{\pi}_{A(EA)})} \qquad (3.4)$$

$$RE(\hat{\pi}_{B(EA)}, \hat{\pi}_{B(SM)}) = \dfrac{MSE(\hat{\pi}_{B(SM)})}{MSE(\hat{\pi}_{B(EA)})} \qquad (3.5)$$

$$RE(\hat{\pi}_{AB(EA)}, \hat{\pi}_{AB(SM)}) = \dfrac{MSE(\hat{\pi}_{AB(SM)})}{MSE(\hat{\pi}_{AB(EA)})} \qquad (3.6)$$

## 4. Application

We set $P = 0.6$ and $\lambda = 0.7$ in order to ensure moderate confidentiality and obtain reliable information from respondents. We fixed $\pi_{AB} < \pi_A$, $\pi_{AB} < \pi_B$ and $\pi_A + \pi_B < 0.99$; the value of $\pi_{AB}$ were also fixed at 0.05, 0.1 and 0.2 as proposed by Lee et al (2013) while $\pi_A$ and $\pi_B$ were changed from 0.1 to 0.9 with a step of 0.1. It is important to note that the sample size does not influence the relative efficiency.

The result of the analysis shows that the proposed estimators $\hat{\pi}_{A(EA)}, \hat{\pi}_{B(EA)}$ & $\hat{\pi}_{AB(EA)}$ performed better than Lee et al (2013) simple and crossed model estimators $\hat{\pi}_{A(SM)}, \hat{\pi}_{B(SM)}$ & $\hat{\pi}_{AB(SM)}$ and $\hat{\pi}_{A(CM)}, \hat{\pi}_{B(CM)}$ & $\hat{\pi}_{AB(CM)}$ respectively under the conditions stated above. The summary is presented in the tables below.

Table 1: Relative efficiency of the proposed estimators with respect to Lee et al (2013) simple model estimators

| $\pi_A$ | $\pi_B$ | $RE(\hat{\pi}_{A(EA)}, \hat{\pi}_{A(SM)})$ | $RE(\hat{\pi}_{B(EA)}, \hat{\pi}_{B(SM)})$ | $\pi_{AB}$ | $RE(\hat{\pi}_{AB(EA)}, \hat{\pi}_{AB(SM)})$ | $\pi_{AB}$ | $RE(\hat{\pi}_{AB(EA)}, \hat{\pi}_{AB(SM)})$ | $\pi_{AB}$ | $RE(\hat{\pi}_{AB(EA)}, \hat{\pi}_{AB(SM)})$ |
|---|---|---|---|---|---|---|---|---|---|
| 0.1 | 0.1 | 24.52 | 6.02 | 0.05 | 25.07 | 0.1 | 25.05 | 0.2 | 26.08 |
| 0.1 | 0.2 | 24.52 | 5.98 | 0.05 | 24.15 | 0.1 | 24.13 | 0.2 | 25.02 |
| 0.1 | 0.3 | 24.52 | 6.09 | 0.05 | 23.39 | 0.1 | 23.38 | 0.2 | 24.17 |
| 0.1 | 0.4 | 24.52 | 6.37 | 0.05 | 22.75 | 0.1 | 22.75 | 0.2 | 23.46 |
| 0.1 | 0.5 | 24.52 | 6.87 | 0.05 | 22.22 | 0.1 | 22.21 | 0.2 | 22.85 |
| 0.1 | 0.6 | 24.52 | 7.70 | 0.05 | 21.76 | 0.1 | 21.76 | 0.2 | 22.34 |
| 0.1 | 0.7 | 24.52 | 9.18 | 0.05 | 21.36 | 0.1 | 21.36 | 0.2 | 21.90 |
| 0.1 | 0.8 | 24.52 | 12.23 | 0.05 | 21.01 | 0.1 | 21.01 | 0.2 | 21.51 |
| 0.2 | 0.1 | 24.68 | 6.02 | 0.05 | 24.44 | 0.1 | 24.42 | 0.2 | 25.39 |
| 0.2 | 0.2 | 24.68 | 5.98 | 0.05 | 23.61 | 0.1 | 23.60 | 0.2 | 24.44 |
| 0.2 | 0.3 | 24.68 | 6.09 | 0.05 | 22.92 | 0.1 | 22.91 | 0.2 | 23.67 |
| 0.2 | 0.4 | 24.68 | 6.37 | 0.05 | 22.35 | 0.1 | 22.34 | 0.2 | 23.02 |
| 0.2 | 0.5 | 24.68 | 6.87 | 0.05 | 21.85 | 0.1 | 21.85 | 0.2 | 22.47 |
| 0.2 | 0.6 | 24.68 | 7.70 | 0.05 | 21.43 | 0.1 | 21.43 | 0.2 | 21.99 |
| 0.2 | 0.7 | 24.68 | 9.18 | 0.05 | 21.06 | 0.1 | 21.06 | 0.2 | 21.58 |
| 0.3 | 0.1 | 25.49 | 6.02 | 0.05 | 23.86 | 0.1 | 23.84 | 0.2 | 24.75 |
| 0.3 | 0.2 | 25.49 | 5.98 | 0.05 | 23.11 | 0.1 | 23.10 | 0.2 | 23.90 |
| 0.3 | 0.3 | 25.49 | 6.09 | 0.05 | 22.49 | 0.1 | 22.48 | 0.2 | 23.20 |
| 0.3 | 0.4 | 25.49 | 6.37 | 0.05 | 21.96 | 0.1 | 21.96 | 0.2 | 22.60 |
| 0.3 | 0.5 | 25.49 | 6.87 | 0.05 | 21.51 | 0.1 | 21.51 | 0.2 | 22.10 |
| 0.3 | 0.6 | 25.49 | 7.70 | 0.05 | 21.12 | 0.1 | 21.12 | 0.2 | 21.66 |
| 0.4 | 0.1 | 27.08 | 6.02 | 0.05 | 23.32 | 0.1 | 23.31 | 0.2 | 24.17 |
| 0.4 | 0.2 | 27.08 | 5.98 | 0.05 | 22.64 | 0.1 | 22.64 | 0.2 | 23.40 |
| 0.4 | 0.3 | 27.08 | 6.09 | 0.05 | 22.08 | 0.1 | 22.07 | 0.2 | 22.76 |
| 0.4 | 0.4 | 27.08 | 6.37 | 0.05 | 21.60 | 0.1 | 21.59 | 0.2 | 22.22 |
| 0.4 | 0.5 | 27.08 | 6.87 | 0.05 | 21.19 | 0.1 | 21.18 | 0.2 | 21.75 |
| 0.5 | 0.1 | 29.76 | 6.02 | 0.05 | 22.82 | 0.1 | 22.81 | 0.2 | 23.62 |
| 0.5 | 0.2 | 29.76 | 5.98 | 0.05 | 22.21 | 0.1 | 22.20 | 0.2 | 22.93 |
| 0.5 | 0.3 | 29.76 | 6.09 | 0.05 | 21.69 | 0.1 | 21.69 | 0.2 | 22.34 |
| 0.5 | 0.4 | 29.76 | 6.37 | 0.05 | 21.25 | 0.1 | 21.25 | 0.2 | 21.85 |
| 0.6 | 0.1 | 34.21 | 6.02 | 0.05 | 22.35 | 0.1 | 22.34 | 0.2 | 23.12 |
| 0.6 | 0.2 | 34.21 | 5.98 | 0.05 | 21.80 | 0.1 | 21.79 | 0.2 | 22.49 |
| 0.6 | 0.3 | 34.21 | 6.09 | 0.05 | 21.33 | 0.1 | 21.33 | 0.2 | 21.96 |
| 0.7 | 0.1 | 42.07 | 6.02 | 0.05 | 21.91 | 0.1 | 21.91 | 0.2 | 22.64 |
| 0.7 | 0.2 | 42.07 | 5.98 | 0.05 | 21.41 | 0.1 | 21.41 | 0.2 | 22.07 |
| 0.8 | 0.1 | 58.33 | 6.02 | 0.05 | 21.51 | 0.1 | 21.50 | 0.2 | 22.20 |

Table 2: Relative efficiency of the proposed estimators with respect to Lee et al (2013) crossed model estimators when $\pi_{AB}$ is fixed at 0.05 and 0.1

| $\pi_A$ | $\pi_B$ | $\pi_{AB}$ | $RE(\hat{\pi}_{A(EA)}, \hat{\pi}_{A(CM)})$ | $RE(\hat{\pi}_{B(EA)}, \hat{\pi}_{B(CM)})$ | $RE(\hat{\pi}_{AB(EA)}, \hat{\pi}_{AB(CM)})$ | $\pi_A$ | $\pi_B$ | $\pi_{AB}$ | $RE(\hat{\pi}_{A(EA)}, \hat{\pi}_{A(CM)})$ | $RE(\hat{\pi}_{B(EA)}, \hat{\pi}_{B(CM)})$ | $RE(\hat{\pi}_{AB(EA)}, \hat{\pi}_{AB(CM)})$ |
|---|---|---|---|---|---|---|---|---|---|---|---|
| 0.1 | 0.1 | 0.05 | 6.4 | 4.6 | 3.0 | 0.1 | 0.1 | 0.1 | 7.1 | 5.0 | 3.5 |
| 0.1 | 0.2 | 0.05 | 5.8 | 4.2 | 2.4 | 0.1 | 0.2 | 0.1 | 6.4 | 4.6 | 2.9 |
| 0.1 | 0.3 | 0.05 | 5.1 | 3.9 | 1.9 | 0.1 | 0.3 | 0.1 | 5.8 | 4.3 | 2.4 |
| 0.1 | 0.4 | 0.05 | 4.4 | 3.6 | 1.6 | 0.1 | 0.4 | 0.1 | 5.1 | 4.1 | 2.0 |
| 0.1 | 0.5 | 0.05 | 3.7 | 3.5 | 1.2 | 0.1 | 0.5 | 0.1 | 4.4 | 3.9 | 1.6 |
| 0.1 | 0.6 | 0.05 | 3.1 | 3.3 | 0.9 | 0.1 | 0.6 | 0.1 | 3.7 | 3.9 | 1.3 |
| 0.1 | 0.7 | 0.05 | 2.4 | 3.2 | 0.7 | 0.1 | 0.7 | 0.1 | 3.1 | 3.9 | 1.0 |
| 0.1 | 0.8 | 0.05 | 1.7 | 3.1 | 0.5 | 0.1 | 0.8 | 0.1 | 2.4 | 4.0 | 0.8 |
| 0.2 | 0.1 | 0.05 | 6.0 | 4.1 | 2.6 | 0.2 | 0.1 | 0.1 | 6.7 | 4.6 | 3.1 |
| 0.2 | 0.2 | 0.05 | 5.4 | 3.7 | 2.1 | 0.2 | 0.2 | 0.1 | 6.0 | 4.2 | 2.5 |
| 0.2 | 0.3 | 0.05 | 4.7 | 3.4 | 1.6 | 0.2 | 0.3 | 0.1 | 5.4 | 3.9 | 2.1 |
| 0.2 | 0.4 | 0.05 | 4.0 | 3.2 | 1.3 | 0.2 | 0.4 | 0.1 | 4.7 | 3.6 | 1.7 |
| 0.2 | 0.5 | 0.05 | 3.3 | 3.0 | 1.0 | 0.2 | 0.5 | 0.1 | 4.0 | 3.5 | 1.4 |
| 0.2 | 0.6 | 0.05 | 2.7 | 2.8 | 0.7 | 0.2 | 0.6 | 0.1 | 3.3 | 3.3 | 1.1 |
| 0.2 | 0.7 | 0.05 | 2.0 | 2.6 | 0.5 | 0.2 | 0.7 | 0.1 | 2.7 | 3.2 | 0.8 |
| 0.3 | 0.1 | 0.05 | 5.7 | 3.6 | 2.2 | 0.3 | 0.1 | 0.1 | 6.4 | 4.1 | 2.7 |
| 0.3 | 0.2 | 0.05 | 5.0 | 3.3 | 1.7 | 0.3 | 0.2 | 0.1 | 5.7 | 3.7 | 2.2 |
| 0.3 | 0.3 | 0.05 | 4.3 | 3.0 | 1.4 | 0.3 | 0.3 | 0.1 | 5.0 | 3.4 | 1.8 |
| 0.3 | 0.4 | 0.05 | 3.6 | 2.8 | 1.0 | 0.3 | 0.4 | 0.1 | 4.3 | 3.2 | 1.4 |
| 0.3 | 0.5 | 0.05 | 2.9 | 2.5 | 0.8 | 0.3 | 0.5 | 0.1 | 3.6 | 3.0 | 1.1 |
| 0.3 | 0.6 | 0.05 | 2.2 | 2.3 | 0.5 | 0.3 | 0.6 | 0.1 | 2.9 | 2.8 | 0.9 |
| 0.4 | 0.1 | 0.05 | 5.4 | 3.2 | 1.8 | 0.4 | 0.1 | 0.1 | 6.1 | 3.6 | 2.3 |
| 0.4 | 0.2 | 0.05 | 4.7 | 2.8 | 1.4 | 0.4 | 0.2 | 0.1 | 5.4 | 3.3 | 1.9 |
| 0.4 | 0.3 | 0.05 | 4.0 | 2.6 | 1.1 | 0.4 | 0.3 | 0.1 | 4.7 | 3.0 | 1.5 |
| 0.4 | 0.4 | 0.05 | 3.2 | 2.3 | 0.8 | 0.4 | 0.4 | 0.1 | 4.0 | 2.8 | 1.2 |
| 0.4 | 0.5 | 0.05 | 2.5 | 2.0 | 0.6 | 0.4 | 0.5 | 0.1 | 3.2 | 2.5 | 0.9 |
| 0.5 | 0.1 | 0.05 | 5.2 | 2.7 | 1.5 | 0.5 | 0.1 | 0.1 | 6.0 | 3.2 | 2.0 |
| 0.5 | 0.2 | 0.05 | 4.4 | 2.4 | 1.1 | 0.5 | 0.2 | 0.1 | 5.2 | 2.8 | 1.6 |
| 0.5 | 0.3 | 0.05 | 3.6 | 2.1 | 0.8 | 0.5 | 0.3 | 0.1 | 4.4 | 2.6 | 1.2 |
| 0.5 | 0.4 | 0.05 | 2.8 | 1.9 | 0.6 | 0.5 | 0.4 | 0.1 | 3.6 | 2.3 | 0.9 |
| 0.6 | 0.1 | 0.05 | 5.0 | 2.2 | 1.2 | 0.6 | 0.1 | 0.1 | 5.9 | 2.7 | 1.7 |
| 0.6 | 0.2 | 0.05 | 4.1 | 2.0 | 0.9 | 0.6 | 0.2 | 0.1 | 5.0 | 2.4 | 1.3 |
| 0.6 | 0.3 | 0.05 | 3.2 | 1.7 | 0.6 | 0.6 | 0.3 | 0.1 | 4.1 | 2.1 | 1.0 |
| 0.7 | 0.1 | 0.05 | 4.8 | 1.8 | 0.9 | 0.7 | 0.1 | 0.1 | 6.0 | 2.2 | 1.4 |
| 0.7 | 0.2 | 0.05 | 3.7 | 1.5 | 0.6 | 0.7 | 0.2 | 0.1 | 4.8 | 2.0 | 1.0 |
| 0.8 | 0.1 | 0.05 | 4.7 | 1.3 | 0.7 | 0.8 | 0.1 | 0.1 | 6.3 | 1.8 | 1.1 |

Table 3: Relative efficiency of the proposed estimators with respect to Lee et al (2013) crossed model estimators when $\pi_{AB}$ is fixed at 0.2

| $\pi_A$ | $\pi_B$ | $\pi_{AB}$ | $RE(\hat{\pi}_{A(EA)}, \hat{\pi}_{A(CM)})$ | $RE(\hat{\pi}_{B(EA)}, \hat{\pi}_{B(CM)})$ | $RE(\hat{\pi}_{AB(EA)}, \hat{\pi}_{AB(CM)})$ |
|---|---|---|---|---|---|
| 0.2 | 0.2 | 0.2 | 7.4 | 5 | 3.5 |
| 0.2 | 0.3 | 0.2 | 6.7 | 4.7 | 3 |
| 0.2 | 0.4 | 0.2 | 6 | 4.5 | 2.5 |
| 0.2 | 0.5 | 0.2 | 5.4 | 4.4 | 2.1 |
| 0.2 | 0.6 | 0.2 | 4.7 | 4.4 | 1.8 |
| 0.3 | 0.2 | 0.2 | 7.1 | 4.6 | 3.1 |
| 0.3 | 0.3 | 0.2 | 6.4 | 4.3 | 2.6 |
| 0.3 | 0.4 | 0.2 | 5.7 | 4.1 | 2.2 |
| 0.3 | 0.5 | 0.2 | 5 | 3.9 | 1.8 |
| 0.4 | 0.2 | 0.2 | 6.9 | 4.2 | 2.8 |
| 0.4 | 0.3 | 0.2 | 6.1 | 3.9 | 2.3 |
| 0.4 | 0.4 | 0.2 | 5.4 | 3.6 | 1.9 |
| 0.5 | 0.2 | 0.2 | 6.8 | 3.7 | 2.4 |
| 0.5 | 0.3 | 0.2 | 6 | 3.4 | 2 |
| 0.6 | 0.2 | 0.2 | 6.8 | 3.3 | 2.1 |

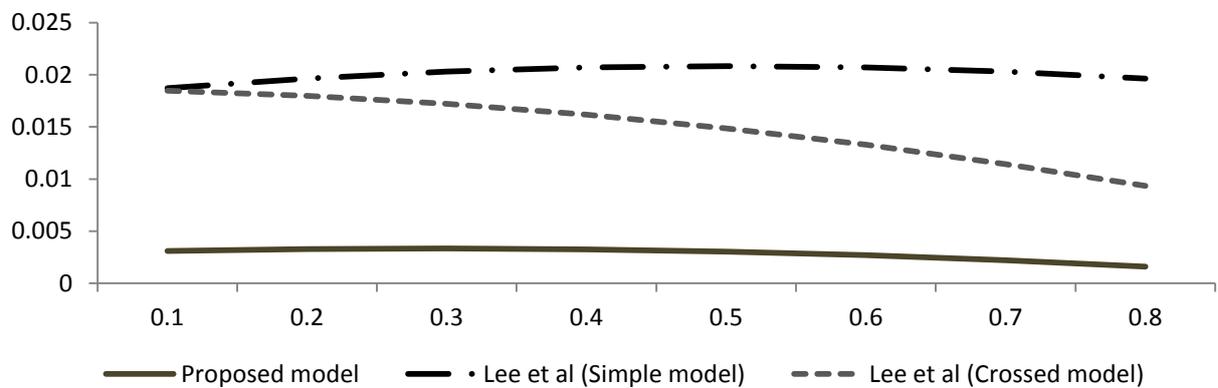

Figure 1: Comparison of the variances of the estimators

## 5. Discussion

In computing relative efficiency of the propose estimators over Lee et al (2013) equations 1.11, 1.12, 1.13, 1.17, 1.18, 1.19, 2.9, 2.10, 2.11, 3.4, 3.5 and 3.6 were used. It was observed that the relative efficiency of $RE(\hat{\pi}_{A(EA)}, \hat{\pi}_{A(SM)})$ and $RE(\hat{\pi}_{B(EA)}, \hat{\pi}_{B(SM)})$ increases with increase in the values of $\pi_A$ and $\pi_B$ respectively. Also, we observed that there is an insignificant drop in $RE(\hat{\pi}_{AB(EA)}, \hat{\pi}_{AB(SM)})$ as $\pi_{AB}$ value increases from 0.05 to 0.1 while a noticeable increase occurs in the relative efficiency, $RE(\hat{\pi}_{AB(EA)}, \hat{\pi}_{AB(SM)})$ as the value of $\pi_{AB}$ increases from 0.1 to 0.2. For the crossed model, it was observed that as the value of $\pi_{AB}$ increases from 0.05 to 0.2 the proposed estimators perform better. This implies that as the proposed method captures more people bearing the joint sensitive characters, the estimators become more efficient.

The crossed model by Lee et al (2013) was adjudged to be better than the simple model which is a special case of Christofides (2005). Also Perri et al (2015) application of crossed model appears encouraging as reported by them but they saw need to re-define it but our simple model is more efficient than them all. This further justifies need for refinement of the crossed model as mentioned by Perri et al (2015).

## 6. Conclusion

We have presented new and more efficient estimators of estimating proportion of people having two related sensitive attributes by extending the work of Mangat (1994). The propose estimator becomes more efficient as the design captures more people possessing the sensitive attribute.